\documentclass[11pt]{article}

\usepackage[margin=1in]{geometry}
\usepackage{amsmath,amssymb,bm}
\usepackage[T1]{fontenc}
\usepackage{graphicx}
\usepackage{caption}
\usepackage{booktabs}
\usepackage{tabularx}
\usepackage{array}
\usepackage{multirow}
\usepackage{enumitem}
\usepackage{microtype}
\usepackage[numbers,square,sort&compress]{natbib}
\usepackage{xcolor}
\usepackage{hyperref}
\usepackage[nameinlink,noabbrev]{cleveref}
\usepackage{placeins}

\hypersetup{
  colorlinks=true,
  linkcolor=blue!55!black,
  citecolor=blue!55!black,
  urlcolor=blue!55!black
}

\graphicspath{{figures/word_revision/}}
\newcommand{\R}{\mathbb{R}}
\newcommand{\RMSE}{\operatorname{RMSE}}
\newcommand{\MAE}{\operatorname{MAE}}

\title{Geometry-Aware Smoothing Selection for Noisy Scattered Surface Reconstruction:\\
Evidence from Synthetic Benchmarks and Water Quality Data}
\author{Mirkan Emir Sancak}
\date{\today}

\begin{document}

\maketitle

\begin{abstract}
Sampling geometry can affect both the stability of scattered surface reconstruction and the regularization needed to control noise. This study distinguishes two questions: whether geometry predicts failure of exact multiquadric radial basis function (RBF) interpolation, and whether it explains the smoothing magnitude selected by validation. A synthetic benchmark crossed four response families, four 16-point geometries, three slice levels, five observation regimes, and 24 repeated holdouts, yielding 17,280 method-level evaluations. Clough--Tocher cubic interpolation, exact RBF interpolation, and smoothed RBF reconstruction were compared, followed by a Water Quality Portal pilot with 1,950 complete activities. On the synthetic shared-valid domain, exact RBF had the lowest noise-free mean RMSE (0.0206), whereas smoothed RBF performed best under noise (0.1807 versus 0.2414 for cubic and 0.6106 for exact RBF). Clustered-gap support produced the largest noisy exact-RBF RMSE (1.2706) and a negative-$R^2$ rate of 0.4175. Geometry added little explanatory power for $\log_{10}$ smoothing: the combined model had $R^2$ = 0.4259 compared with 0.4096 for noise alone. By contrast, the supplied method-valid failure analysis reported an increase in McFadden pseudo-$R^2$ from 0.180 to 0.266 when geometry was added to noise, with a significant geometry-block test under slice-task-clustered inference. In the environmental pilot, smoothed RBF achieved mean RMSE 0.9877 with no negative-$R^2$ failures, while exact RBF either produced no valid prediction or failed on the finite predictions. Repeated predictor coordinates with conflicting responses and strongly unequal coordinate scales made exact fitting unsuitable. All environmental fits selected the largest tested smoothing value, preventing a geometry--smoothing correlation analysis. The findings support geometry-aware failure screening within the studied designs, but not a universal monotone rule for smoothing magnitude. Independent multisite validation, joint scale--smoothing tuning, and direct conditioning diagnostics are needed before an adaptive policy can be claimed.
\end{abstract}


\section{Introduction}
\label{sec:1}

Irregularly located observations are the norm rather than the exception in environmental monitoring, geoscience, remote sensing, and process engineering, where measurement locations are fixed by access, cost, monitoring infrastructure, experimental design, or historical practice rather than by an ideal numerical layout \citep{ganesan2004,trewin2010,smith2017,li2014}. Reconstructing a continuous response from such observations is therefore a recurring problem across numerical approximation, environmental analysis, and scientific visualization \citep{amidror2002,fussen2003,luo2024,si2026}. As a result, datasets differ not only in sample size but in spacing, clustering, coverage, and configuration, all of which influence reconstruction accuracy and uncertainty \citep{aguilar2005,li2011,diefenbach2025}. Comparative environmental studies show that sampling design can affect interpolation performance independently of density \citep{li2011}. Irregular sampling is not inherently harmful: in sparsity-promoting reconstruction some nonregular designs outperform regular subsampling, precisely because random or jittered layouts convert coherent aliasing into incoherent noise while bounding the maximum gap \citep{hennenfent2008,herrmann2010}. The central issue is therefore not irregularity itself, but whether the observations provide balanced support over the domain or create geometrically challenging structures such as clusters, large gaps, anisotropy, or boundary concentration.

Radial basis function methods provide a meshfree framework for reconstructing smooth surfaces from scattered observations. The multiquadric construction introduced by \citet{hardy1971} was placed on a rigorous solvability foundation by \citet{micchelli1986}. Subsequent theory connects approximation quality to the geometry of the data sites. For $X\subset\Omega$, the fill distance measures the largest distance from a point in the domain to its nearest observation, while the separation distance conventionally equals half the minimum distance between distinct sites. Their ratio, the mesh ratio, summarizes nonuniformity \citep{wendland2005,narcowich2005}. Under their respective assumptions, scattered-data approximation results express error or stability bounds in terms of fill distance and site distribution \citep{vongolitschek2001,demarchi2010,hangelbroek2012}. Related moving least-squares constructions extend meshfree approximation to divergence-free and curl-free fields \citep{mirzaei2026}. RBF accuracy and matrix conditioning also depend jointly on kernel scale and separation, so a geometrically uneven design may be sensitive to response perturbations without being inaccurate for every clean target \citep{schaback1995}. Recent regression work links reduced training-set fill distance to smaller worst-case prediction error and improved Gaussian-kernel stability \citep{climaco2024}. Two datasets with equal sample size can therefore present substantially different reconstruction problems.

The connection between geometry and approximation does not itself specify how to choose smoothing. Diagonal regularization is related to ridge regression \citep{hoerl1970} and penalized-spline smoothing \citep{wahba1990}, while generalized cross-validation and kernel-specific validation schemes provide established parameter-selection principles \citep{golub1979,rippa1999,fasshauer2007}. For noisy spherical data, \citet{hesse2017} derive error bounds for discrepancy-based and a priori smoothing strategies. \citet{wendlandrieger2005} connect approximate interpolation and smoothing selection through conditions involving fill distance. Adaptive kernel-ridge selection likewise seeks to match regularization to estimation difficulty \citep{lin2024}. These results motivate geometry-aware analysis, but they do not establish a universal monotone relationship between an empirical gap descriptor and the smoothing value chosen by a finite validation grid.

A complementary line of work addresses heterogeneous support through local or adaptive reconstruction. Local penalized least-squares RBF methods restrict the available data and error analysis to a subdomain \citep{hesse2021}, stabilized greedy algorithms adapt basis-site selection \citep{wenzel2021}, and variably scaled or anisotropic kernels alter the effective representation across the domain \citep{bozzini2015,demarchi2020,rigutto2025}. Together, these approaches show that adaptation can operate through data selection, kernel scale, locality, or a smoothing penalty. The narrower question pursued here is whether a small set of global geometric descriptors provides information about validation-selected smoothing or exact-interpolation failure after empirical noise is accounted for.

The present work builds on the exact-interpolation benchmark of \citet{sancak2026}, which compared cubic and multiquadric surfaces under a shared protocol and found that exact fitting overfits noisy observations, with cubic interpolation comparatively more stable. Here, validation-selected smoothing is added as a third method, and the emphasis shifts from method ranking to the circumstances in which exact fitting becomes unsafe. The first hypothesis, H1, is that smoothing improves predictive robustness under noisy observations. Exact interpolation reproduces measurement error as well as signal, whereas a positive penalty relaxes node matching and may reduce variability at the cost of bias \citep{hastie2009}.

The second hypothesis concerns sampling geometry and has two distinct forms. H2a states that difficult geometry raises exact-interpolation failure risk and provides information about that risk beyond empirical noise. Closely spaced observations and large unsupported regions can coexist, leaving a global fit sensitive to perturbations when it must bridge weakly constrained parts of the domain \citep{schaback1995}. H2b states that, after accounting for noise, larger gaps are associated with stronger validation-selected smoothing. This directional claim is tested separately: a geometry may benefit substantially from regularization without requiring the largest numerical value of a global penalty.

The third hypothesis, H3, is that regularization improves the stability of surface-derived decisions, represented here by estimated optimum location. An optimum depends on the ranking and shape of the fitted surface rather than on average error alone. A small number of spurious peaks can therefore move the estimated optimum substantially, even when most predictions are reasonable. Smoothing may suppress such peaks and improve resampling consistency, although consistency does not by itself establish accuracy when the true optimum is unknown.

The contribution is deliberately two-tiered. The synthetic experiment manipulates geometry and noise under controlled conditions, isolating failure mechanisms, while the Water Quality Portal (WQP) pilot then tests whether the same workflow remains useful in a realistic environmental feature space. The real-data study is not presented as a definitive field validation of a geometry-to-smoothing law; it identifies which parts of the synthetic argument transfer, which do not, and what design changes a decisive environmental test would require.

\section{Materials and Methods}
\label{sec:2}

\subsection{Overall study design}
\label{sec:2.1}

The analysis comprised a controlled synthetic benchmark and a real environmental pilot, both using a slice-wise surface-reconstruction protocol. Within a slice, observations were pairs $\bm{x}_i=(x_{i1},x_{i2})\in\mathbb{R}^2$ with scalar responses $y_i$. Three methods were compared: Clough--Tocher piecewise cubic interpolation (denoted cubic); exact multiquadric RBF interpolation with smoothing \ensuremath{\lambda} = 0 (denoted rbf\_exact); and multiquadric RBF reconstruction with \ensuremath{\lambda} selected by inner validation (denoted rbf\_smooth). The implementation used the scientific Python stack, combining NumPy for array computation \citep{harris2020}, pandas for tabular handling \citep{mckinney2010}, Matplotlib for figures \citep{hunter2007}, and SciPy's CloughTocher2DInterpolator and RBFInterpolator routines \citep{virtanen2020}, with a fixed random seed of 42 to make every reported quantity reproducible.

\subsection{Interpolation models}
\label{sec:2.2}

The cubic comparator constructs a Delaunay triangulation of the training sites using the Qhull library \citep{barber1996} and fits a piecewise cubic, continuously differentiable Bézier surface over that triangulation through a Clough--Tocher macro-element scheme \citep{alfeld1984,farin1986}, with the nodal gradients estimated by the minimum-norm-network method so that the surface curvature is approximately minimized \citep{nielson1983}. Its local polynomial structure suits scattered two-dimensional data, but evaluation is restricted to the convex hull of the training nodes, so predictions outside the hull are undefined; this property motivates the shared-valid-domain comparison described below.

For training sites $X=\{\bm{x}_1,\ldots,\bm{x}_n\}$, the fitted RBF surface has the form

\begin{equation}
\widehat f(\bm{x})=\sum_{j=1}^{n}a_j\phi_{\varepsilon}(\lVert\bm{x}-\bm{x}_j\rVert_2)+p(\bm{x}),
\label{eq:word-16}
\end{equation}

where $\phi_{\varepsilon}$ is the multiquadric kernel \citep{hardy1971,micchelli1986} and p is the default degree-zero polynomial term required by the SciPy formulation. With SciPy's sign convention the unscaled kernel is $-\sqrt{1+r^2}$, with input distance scaled by the shape parameter \ensuremath{\varepsilon}. The coefficients solve the regularized block system

\begin{equation}
\begin{aligned}(K+\lambda I)\bm{a}+P\bm{b}&=\bm{y},\\ P^{\mathsf T}\bm{a}&=\bm{0}.\end{aligned}
\label{eq:word-18}
\end{equation}

where $K_{ij}=\phi_{\varepsilon}(\lVert\bm{x}_i-\bm{x}_j\rVert_2)$. Setting \ensuremath{\lambda} = 0 imposes exact fitting when coordinates are distinct and the augmented system is solvable. Positive \ensuremath{\lambda} relaxes node matching through a diagonal penalty. This has a ridge-type interpretation on the polynomial-constrained RBF coefficient space \citep{hoerl1970,wahba1990}. Multiquadric conditioning depends jointly on shape, smoothing, and site distribution \citep{schaback1995,fornberg2004}. The present implementation uses the median positive pairwise training distance as \ensuremath{\varepsilon}; this is an implementation-specific heuristic, not a jointly optimized shape parameter. The smoothing candidate set was

\begin{equation}
\Lambda=\{10^{-6},10^{-4},10^{-3},10^{-2},0.05,0.10,0.25,0.50\}.
\label{eq:word-20}
\end{equation}

Each candidate was scored over five randomized inner holdouts. In the synthetic benchmark, the outer training set contained 11 points; each inner split fitted eight points and validated on the remaining three. Validation used the observed responses, whereas the outer synthetic test used the clean surface truth. The candidate with the smallest mean validation RMSE was selected. This follows the general principle of validation-based RBF parameter choice \citep{rippa1999,fasshauer2007}, but the implemented procedure is repeated holdout validation, not generalized cross-validation. Only the diagonal smoothing parameter was tuned; the shape heuristic was reapplied to the fitting sites.

\subsection{Synthetic benchmark}
\label{sec:2.3}

The synthetic domain was [1, 2] \ensuremath{\times} [0.5, 1.5], with a third variable fixed at slice levels $x_3\in\{2,3,4\}$, and four response families were chosen to represent different surface structures (\Cref{tab:1}). Each geometry contained 16 sites. The families range from a smooth quadratic to a multi-extremum surface, so that the benchmark spans targets that are easy and hard for a global interpolant to reproduce.

\begin{table}[htbp]
\centering
\small
\caption{Synthetic response families.}
\label{tab:1}
\begin{tabularx}{\textwidth}{@{}lX@{}}
\toprule

Family & Response function f($x_{1}$, $x_{2}$, $x_{3}$) \\

\midrule

Smooth quadratic & $x_1^2+0.8x_2+\sin x_3$ \\

Interaction cubic & $x_1x_2+x_2^3-0.3\cos x_3$ \\

Anisotropic wave & $\cos(1.2x_1)+x_2x_3+0.4\sin(3x_2)$ \\

Multiple extrema & $0.6\sin(3.2x_1)+0.7\cos(2.8x_2)+0.25x_1x_2+0.15x_3$ \\

\bottomrule
\end{tabularx}
\end{table}

Four support patterns were evaluated to represent qualitatively different sampling designs: a 4\ensuremath{\times}4 regular grid, a Gaussian-jittered grid clipped to the domain, two separated eight-point clusters producing an interior coverage gap, and a boundary-dominant pattern. In the current implementation the 16 retained boundary-dominant nodes all lie on the perimeter, making this a deliberately severe boundary-only design rather than a mild boundary bias. These designs echo the recurring finding in the interpolation literature that configuration, not only density, governs accuracy and stability \citep{aguilar2005,li2011}.

The sampling configurations were summarized by geometric descriptors drawn from scattered-data approximation theory \citep{wendland2005,narcowich2005}. Letting $d_i$ denote the nearest-neighbour distance of site $x_{i}$ and letting G be the dense 80\ensuremath{\times}80 reference grid, the empirical fill distance was defined over the grid as

\begin{equation}
h_X=\max_{\bm{g}\in G}\min_{\bm{x}_i\in X}\lVert\bm{g}-\bm{x}_i\rVert_2.
\label{eq:word-28}
\end{equation}

The gap ratio was defined as the 95th percentile of grid-to-site distance divided by the median nearest-neighbour distance, and the remaining descriptors were the mean, coefficient of variation, minimum, and maximum of $d_i$, the convex-hull area divided by the unit-area domain, the fraction of nodes within 0.08 units of the boundary, and the covariance anisotropy ratio $\lambda_{\max}/\lambda_{\min}$. The average profiles in \Cref{tab:2} confirm that the designed patterns differ substantially: the clustered-gap support has the largest fill distance, gap ratio, and anisotropy, whereas the boundary-dominant support has complete hull area but no interior nodes. Because the grid and boundary-dominant patterns are deterministic while the jittered and clustered patterns are regenerated per slice task, the benchmark contains 26 distinct geometry-descriptor configurations that cluster into four macro-classes; the descriptors are strongly correlated by construction, which limits the power of any continuous regression of \ensuremath{\lambda} on a single descriptor, so the failure analysis in \Cref{sec:3.4} treats geometry as a block rather than attributing effects to individual descriptors.

\begin{table}[htbp]
\centering
\small
\caption{Mean geometric profiles of the synthetic support patterns. NN denotes nearest neighbour and CV denotes coefficient of variation.}
\label{tab:2}
\begin{tabularx}{\textwidth}{@{}Xrrrrrr@{}}
\toprule

Geometry & Mean NN & NN CV & Fill dist. & Gap ratio & Hull ratio & Anisotropy \\

\midrule

Grid & 0.333 & 0.000 & 0.233 & 0.598 & 1.000 & 1.000 \\

Jittered & 0.283 & 0.100 & 0.246 & 0.697 & 0.968 & 1.059 \\

Clustered gap & 0.091 & 0.617 & 0.622 & 6.380 & 0.363 & 12.499 \\

Boundary dominant & 0.250 & 0.000 & 0.494 & 1.572 & 1.000 & 1.000 \\

\bottomrule
\end{tabularx}
\end{table}

Five observation regimes were applied to the exact surface values to represent a range of realistic error structures. The noise-free regime kept $y_i=f(\bm{x}_i)$; the low-Gaussian regime added N(0, 0.$10^{2}$) error; the medium-Gaussian regime added N(0, 0.$35^{2}$); the heteroskedastic regime added Gaussian error whose standard deviation rose roughly linearly from 0.08 to 0.38 across the domain; and the outlier-contaminated regime added N(0, 0.$18^{2}$) together with an extra N(0, 1.$25^{2}$) perturbation on one eighth of the nodes. For every generated slice, the realized noise was summarized by its standard deviation, mean absolute magnitude, an RMSE-scale proxy, and an outlier fraction based on 2.5 robust standard deviations, so that selected smoothing could later be related to empirical rather than nominal noise.

Crossing four functions, four geometries, three slice levels, and five regimes produced 240 slice tasks. Each task was evaluated with 24 randomized 70/30 holdouts, yielding 11 training and five test nodes per split. Repeated holdouts describe sensitivity to the train--test partition \citep{kohavi1995}; they are not independent realizations of the surface, support, or noise. Models were fitted to noisy training observations and evaluated against the clean test truth. Three methods per split produced 240 \ensuremath{\times} 24 \ensuremath{\times} 3 = 17,280 method-level evaluations.

\subsection{Water Quality Portal pilot}
\label{sec:2.4}

The WQP web service was queried for stream activities in Prince George's County, Maryland (county code US:24:033), requesting January 1, 2020 through December 31, 2025 and four characteristics: dissolved oxygen, water temperature, pH, and specific conductance. The WQP integrates observations contributed by many monitoring organizations and provides programmatic access to standardized records \citep{read2017,wqp2021}. Although the query requested data through the end of 2025, the cached extract used here contained complete activities only from January 6, 2020 through December 10, 2024. Measurements were retained in mg/L for dissolved oxygen, degrees Celsius for temperature, and the reported pH scale, and the conductance units uS/cm @25C, uS/cm, and umho/cm were treated as equivalent for this pilot. Records were pivoted to one row per activity, monitoring location, and date, averaging repeated within-activity values, and complete-case filtering produced 1,950 activities from 23 monitoring locations.

Dissolved oxygen was the response, temperature and pH formed the two reconstruction coordinates, and conductance defined three approximately equal-frequency slices (\Cref{tab:5}). The geometry in this experiment is therefore geometry in environmental predictor space rather than geographic map geometry, an increasingly common framing as data-driven models are used to relate water-quality variables at large scales \citep{zhi2021}. Coordinates were used in their original units, a choice revisited as a limitation because distance-based descriptors then depend on variable scale. Within each slice, 40 randomized 70/30 holdouts were performed with the same interpolation code, RBF scale rule, and candidate set as the synthetic benchmark, predictions were evaluated against held-out observed dissolved oxygen, and the fitted maximum over an 80\ensuremath{\times}80 temperature--pH grid was recorded to assess resampling stability; because no true environmental optimum exists, only dispersion across splits can be reported.

\subsection{Evaluation metrics and comparison domains}
\label{sec:2.5}

For valid predictions $\widehat y_i$ the principal metrics were the root-mean-square error (RMSE), the mean absolute error (MAE), and the coefficient of determination ($R^2$):

\begin{equation}
\begin{aligned}\RMSE&=\sqrt{\frac{1}{m}\sum_{i=1}^m(y_i-\widehat y_i)^2},\\\MAE&=\frac{1}{m}\sum_{i=1}^m|y_i-\widehat y_i|,\\R^2&=1-\frac{\sum_i(y_i-\widehat y_i)^2}{\sum_i(y_i-\bar y)^2}.\end{aligned}
\label{eq:word-39}
\end{equation}

RMSE and MAE were retained because they respond differently to error dispersion: RMSE is more sensitive to large deviations, whereas MAE summarizes absolute error without squaring it \citep{willmott2005}. The 95th percentile and maximum absolute errors describe the tails. A split with $R^2$ < 0 was classified as a predictive failure because it performed worse than the test-set mean baseline. Fits returning no finite prediction were tracked separately; they must not be counted as successful fits merely because $R^2$ is undefined.

Two evaluation domains were retained. The method-valid domain uses every finite prediction returned by a method. The shared-valid domain uses only test points with finite predictions from all three methods in the same split. Primary synthetic accuracy comparisons use the shared-valid domain, with coverage reported separately. The additional failure-regression analysis uses method-valid $R^2$ unless explicitly identified as a shared-valid sensitivity analysis. In the synthetic experiment, optimum-location error is the Euclidean distance between the true and fitted grid maxima. Cubic optimum error is available only when the fitted surface is finite over the complete reference grid, so its aggregate is based on a restricted subset and is not directly comparable with unrestricted RBF optimum errors.

\subsection{Geometry, smoothing, and failure analysis}
\label{sec:2.6}

For smoothed-RBF runs, the response was z = $\log_{10}(\lambda)$. Pearson correlations describe marginal associations, and exploratory ordinary least squares compares standardized geometry-only, noise-only, and combined predictor blocks. For exact-RBF failure, logistic regression models the binary indicator of method-valid $R^2$ < 0 \citep{hosmer2013}. Model summaries include McFadden pseudo-$R^2$ \citep{mcfadden1974} and area under the ROC curve \citep{fawcett2006}; these are descriptive fitted-sample measures rather than estimates of performance on independent geometries.

The supplied failure-analysis summary reports a joint test of a six-coefficient geometry block with standard errors clustered by slice task \citep{cameron2015}. Aggregation to 240 slice-task means is also reported to reduce dependence arising from repeated holdouts. Neither procedure makes the underlying geometry configurations independent: support is reused across noise regimes, and deterministic designs recur across tasks. These limitations are important for avoiding pseudoreplication \citep{hurlbert1984}. The exact six-descriptor specification, clustering identifiers, and analysis code need to accompany the results for reproduction. Accordingly, the inference is treated as exploratory, and no causal effect or out-of-geometry predictive validity is claimed.

\section{Results}
\label{sec:3}

\subsection{Synthetic reconstruction accuracy and coverage}
\label{sec:3.1}

The mean valid fraction was 0.515 for cubic interpolation and 1.000 for both RBF methods, confirming that full-domain averages alone do not provide a like-for-like comparison and justifying the shared-valid view. The shared-valid results in \Cref{tab:3} show a clear regime change. Under noise-free observations, exact RBF achieved the lowest mean RMSE (0.0206), followed by smoothed RBF (0.0273) and cubic (0.0545), which is consistent with the accuracy of exact interpolation on clean, internally consistent data. Across the four noisy regimes, smoothed RBF was best (0.1807), followed by cubic (0.2414), while exact RBF deteriorated to 0.6106, reproducing at the level of a controlled benchmark the tendency of exact fitting to overfit noisy nodes reported previously \citep{sancak2026}.

\begin{table}[htbp]
\centering
\small
\caption{Synthetic shared-valid-domain results, averaged across geometry--regime summary cells. Negative-$R^2$ rate is the fraction of repeated splits with $R^2$ < 0. Cubic optimum error is selectively observed (available only where the fitted surface is finite over the full reference grid).}
\label{tab:3}
\begin{tabularx}{\textwidth}{@{}Xrrrr@{}}
\toprule

Observation set & Method & RMSE & Neg-$R^2$ rate & Optimum error \\

\midrule

Noise free & Cubic & 0.0545 & 0.0530 & 0.0000 \\

Noise free & Exact RBF & 0.0206 & 0.0061 & 0.0642 \\

Noise free & Smoothed RBF & 0.0273 & 0.0104 & 0.0495 \\

Noisy regimes & Cubic & 0.2414 & 0.1975 & 0.2175 \\

Noisy regimes & Exact RBF & 0.6106 & 0.3164 & 0.4158 \\

Noisy regimes & Smoothed RBF & 0.1807 & 0.1589 & 0.1982 \\

\bottomrule
\end{tabularx}
\end{table}

\begin{figure}[htbp]
\centering
\includegraphics[width=\linewidth,height=0.77\textheight,keepaspectratio]{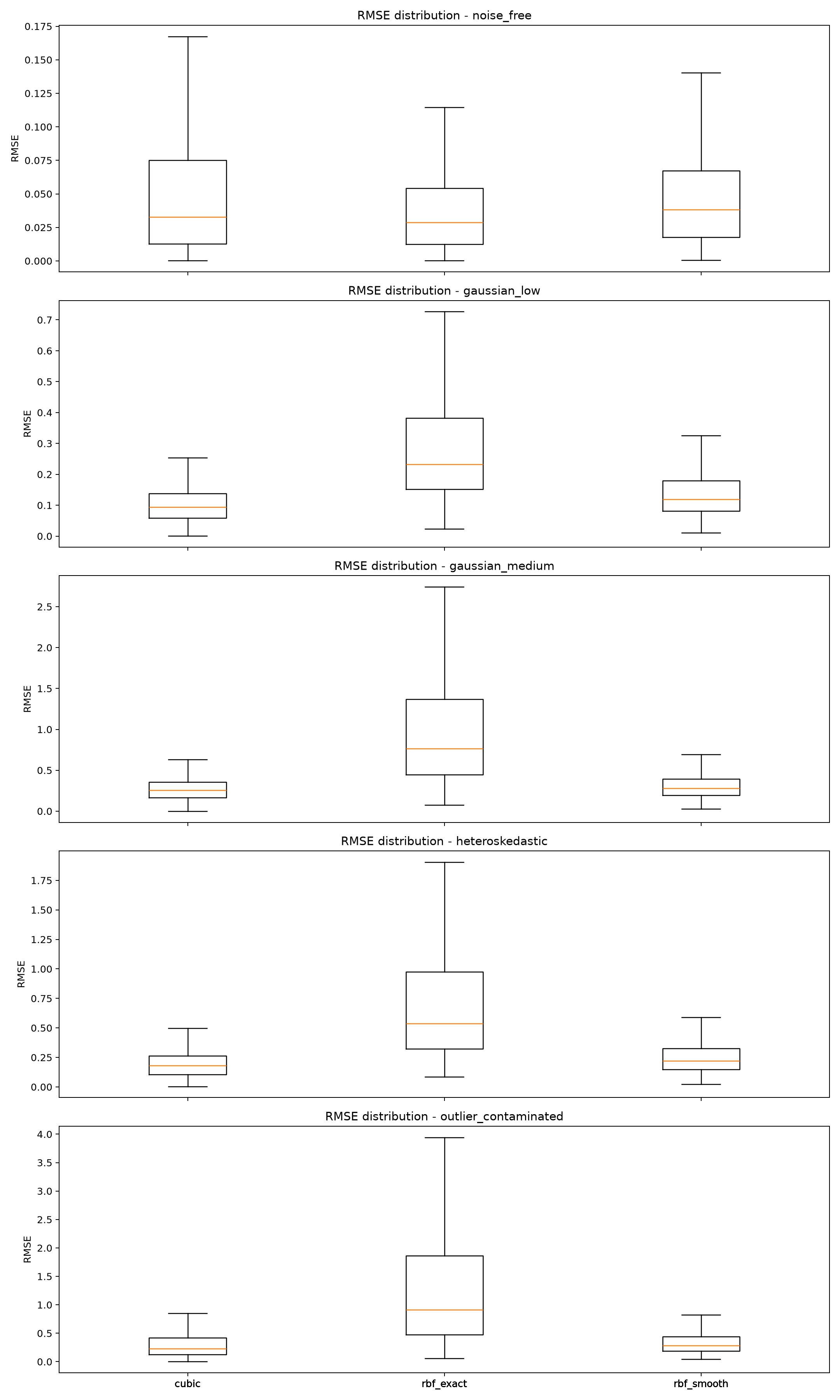}
\caption{Run-level RMSE distributions by observation regime and method. Exact RBF develops a pronounced upper tail as noise increases. Distributions use each method's finite predictions; shared-valid means are reported in \Cref{tab:3}.}
\label{fig:1}
\end{figure}

\begin{figure}[htbp]
\centering
\includegraphics[width=\linewidth,height=0.77\textheight,keepaspectratio]{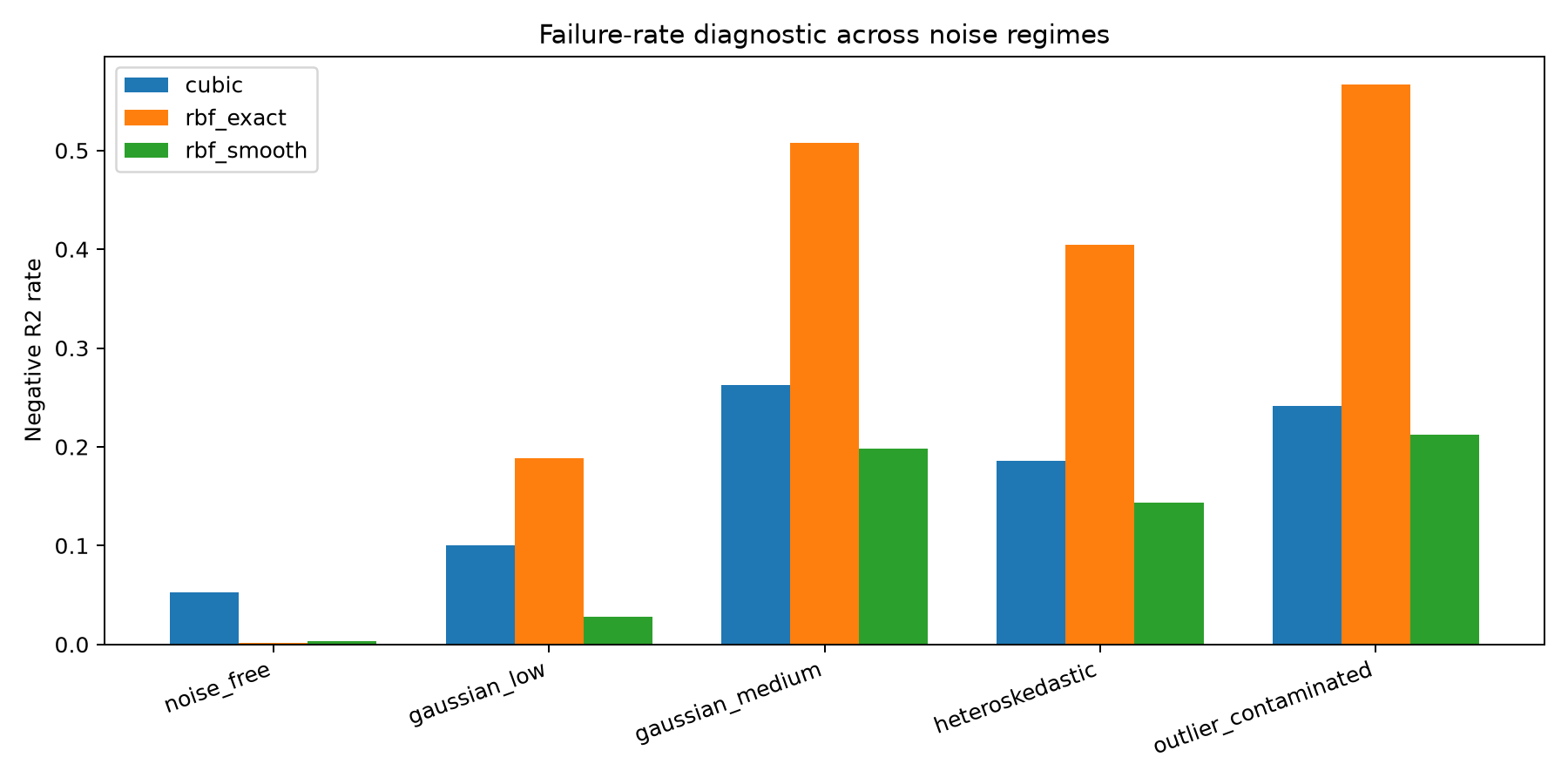}
\caption{Negative-$R^2$ failure rates across observation regimes using the method-valid summaries. Failure risk rises sharply for exact RBF under medium Gaussian, heteroskedastic, and outlier-contaminated noise.}
\label{fig:2}
\end{figure}

Across noisy regimes, smoothing reduced the shared-valid negative-$R^2$ rate from 0.3164 to 0.1589 relative to exact RBF, a relative reduction of approximately 50\%. This is consistent with, but does not independently measure, a favorable bias--variance trade-off \citep{hastie2009}. \Cref{fig:1} and \Cref{fig:2} show method-valid error distributions and failure rates, respectively; their evaluation domains differ from the shared-valid summaries in \Cref{tab:3}. Mean $R^2$ was highly sensitive to splits with very small test-response variance, particularly after restricting the already small test set to the shared domain. Negative-$R^2$ rates are therefore reported alongside RMSE rather than relying on mean $R^2$ alone.

\subsection{Geometry controls where exact interpolation fails}
\label{sec:3.2}

The clustered-gap pattern was the decisive stress test (\Cref{tab:4}). Across noisy regimes, exact RBF reached shared RMSE 1.2706 and a negative-$R^2$ rate of 0.4175, substantially worse than for grid, jittered, or boundary-dominant support, whereas smoothed RBF reduced clustered-gap RMSE to 0.1500, an 88\% reduction relative to exact RBF. Difficult geometry thus amplified the cost of exact fitting even though the same geometry did not always select the largest numerical smoothing value, a pattern that is coherent with conditioning theory, in which poorly separated and unevenly distributed sites inflate the condition number of the kernel matrix and magnify the effect of response perturbations \citep{schaback1995,demarchi2010}.

\begin{table}[htbp]
\centering
\small
\caption{Synthetic noisy-regime results by geometry. RMSE and failure rates use the shared-valid domain. Mean \ensuremath{\lambda} applies only to smoothed RBF.}
\label{tab:4}
\begin{tabularx}{\textwidth}{@{}Xrrrr@{}}
\toprule

Geometry & Method & RMSE & Neg-$R^2$ rate & Mean \ensuremath{\lambda} \\

\midrule

Grid & Cubic & 0.2094 & 0.1536 & --- \\

Grid & Exact RBF & 0.3780 & 0.2743 & 0 \\

Grid & Smoothed RBF & 0.1856 & 0.1267 & 0.0482 \\

Jittered & Cubic & 0.2024 & 0.1658 & --- \\

Jittered & Exact RBF & 0.4120 & 0.3064 & 0 \\

Jittered & Smoothed RBF & 0.1848 & 0.1450 & 0.0430 \\

Clustered gap & Cubic & 0.3114 & 0.2760 & --- \\

Clustered gap & Exact RBF & 1.2706 & 0.4175 & 0 \\

Clustered gap & Smoothed RBF & 0.1500 & 0.1866 & 0.0661 \\

Boundary dominant & Cubic & 0.2423 & 0.1944 & --- \\

Boundary dominant & Exact RBF & 0.3817 & 0.2674 & 0 \\

Boundary dominant & Smoothed RBF & 0.2025 & 0.1771 & 0.0861 \\

\bottomrule
\end{tabularx}
\end{table}

\begin{figure}[htbp]
\centering
\includegraphics[width=\linewidth,height=0.77\textheight,keepaspectratio]{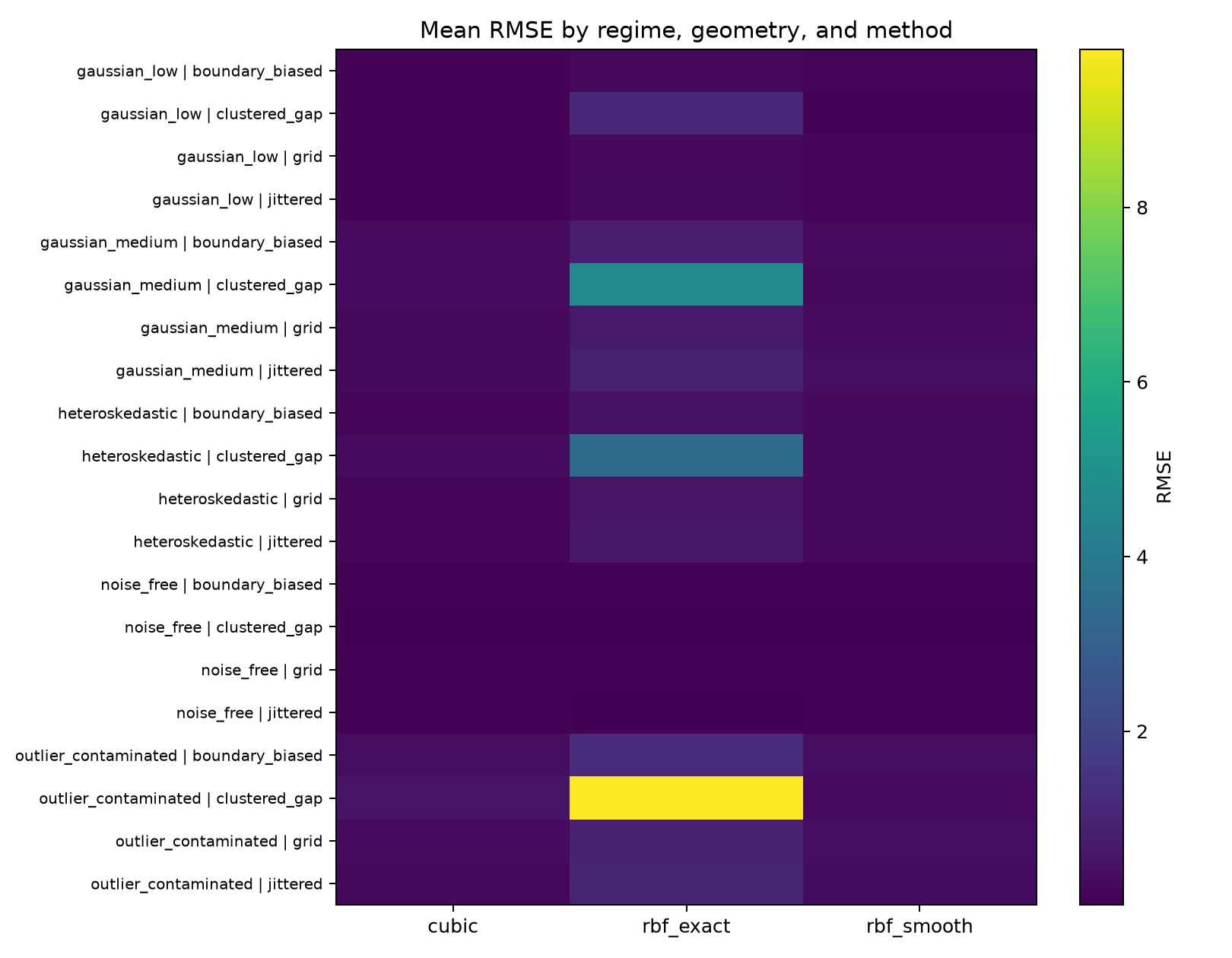}
\caption{Mean RMSE by observation regime, geometry, and method. The exact-RBF column concentrates its error in the clustered-gap rows, most extremely under outlier contamination.}
\label{fig:3}
\end{figure}

The geometry effect was conditional on noise. In the clean clustered-gap case, exact-RBF shared RMSE was only 0.0036, showing that a large gap is not intrinsically fatal for smooth deterministic targets, whereas under noise the same geometry let local perturbations propagate through a poorly constrained global fit. The practical interpretation is an interaction in which geometry determines vulnerability while noise activates it. \Cref{fig:3} shows the same structure at the cell level, with the outlier-contaminated clustered-gap cell dominating the exact-RBF column.

\subsection{Selected smoothing is predominantly noise-driven}
\label{sec:3.3}

Mean selected smoothing increased systematically with observation difficulty, rising from 0.0016 for noise-free data to 0.0079 for low Gaussian, 0.0427 for heteroskedastic, 0.0811 for medium Gaussian, and 0.1119 for outlier-contaminated noise. The candidate-frequency distribution agrees, since the smallest candidate $10^{-6}$ was selected in 71.5\% of noise-free splits whereas the largest candidate 0.50 was selected in 16.2\% of outlier-contaminated splits, which is the behaviour expected when a validation criterion tracks the noise level \citep{golub1979,fasshauer2007}.

\begin{figure}[htbp]
\centering
\includegraphics[width=\linewidth,height=0.77\textheight,keepaspectratio]{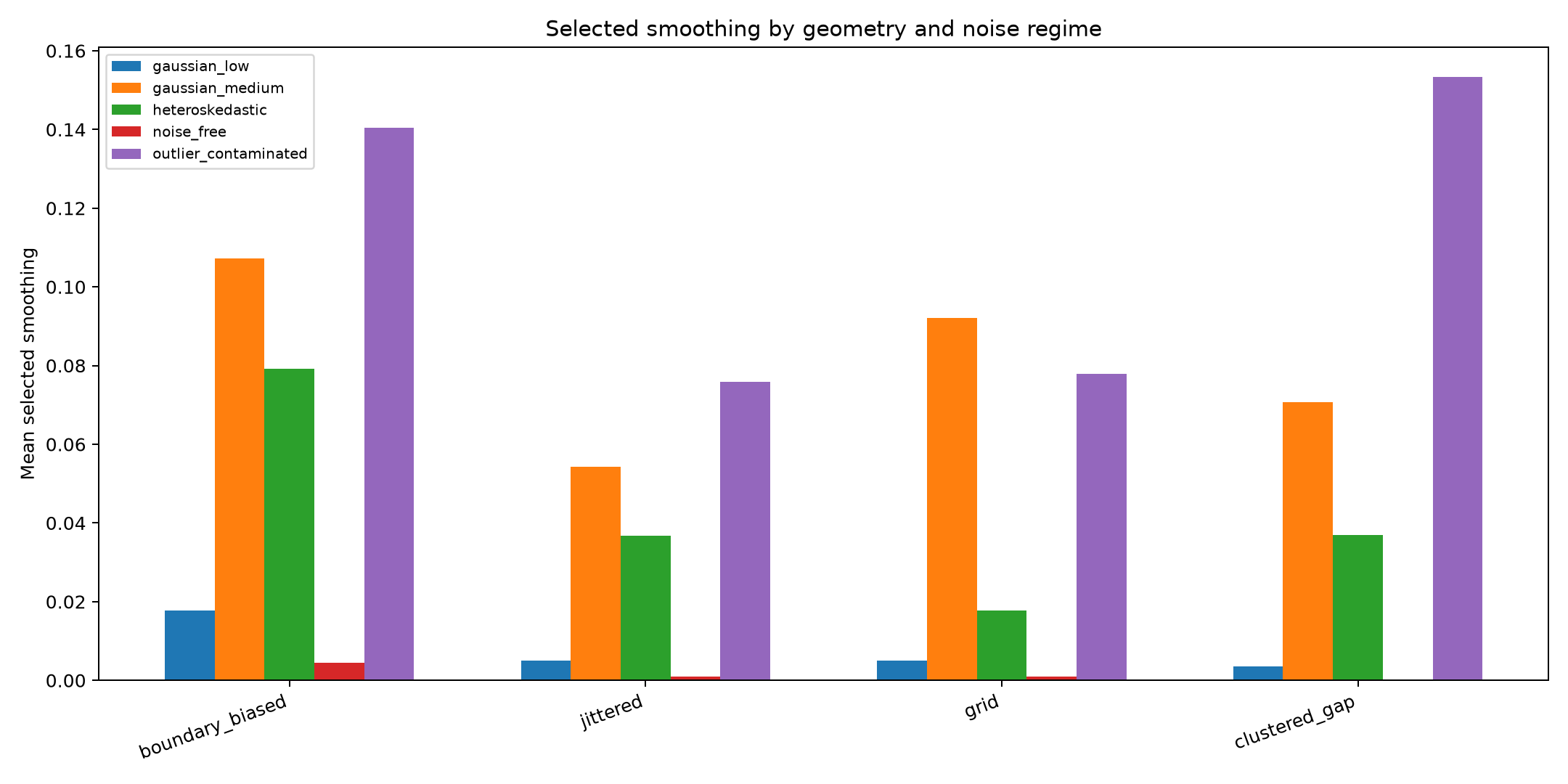}
\caption{Mean selected smoothing by geometry and observation regime. The dominant ordering is by noise regime; geometry changes magnitude within regimes but not monotonically in fill distance or gap ratio.}
\label{fig:4}
\end{figure}

\begin{figure}[htbp]
\centering
\includegraphics[width=\linewidth,height=0.77\textheight,keepaspectratio]{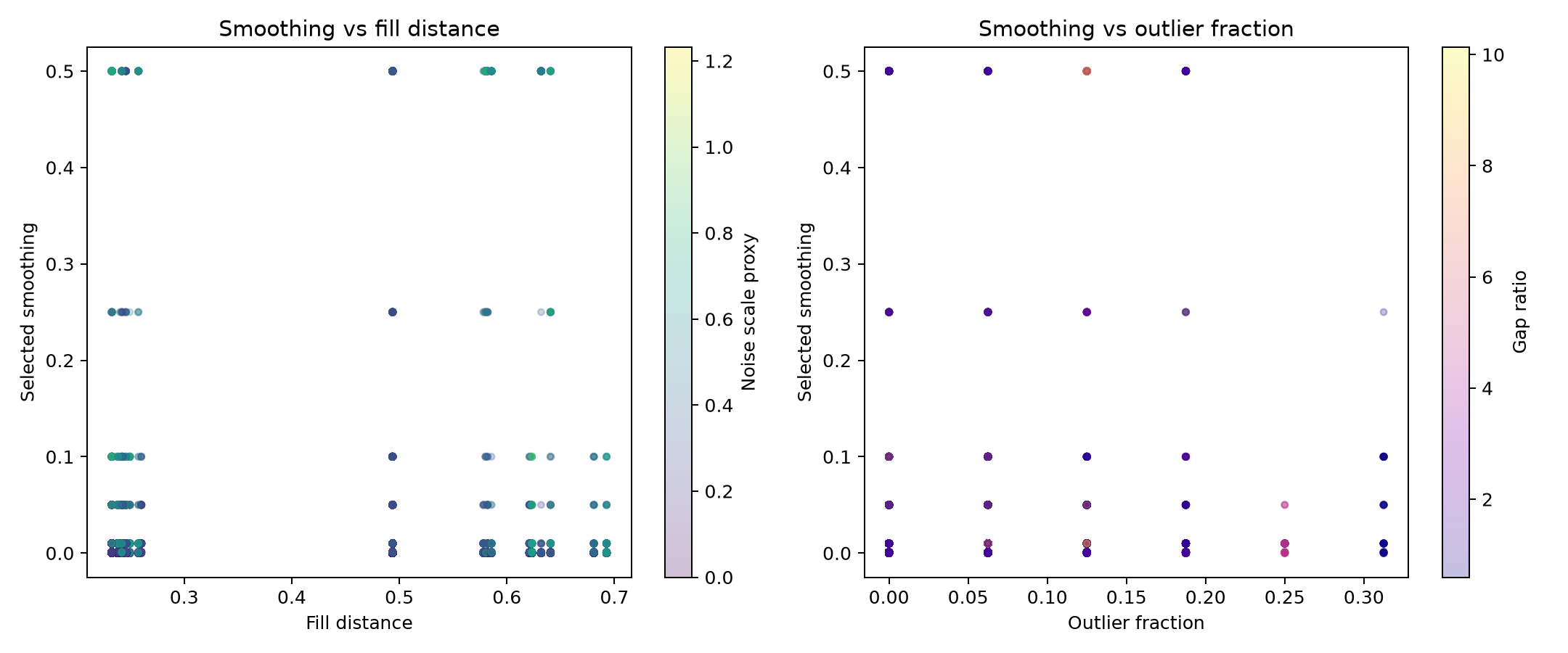}
\caption{Selected smoothing against fill distance and realized outlier fraction. Discrete horizontal bands reflect the finite candidate set; colour encodes a second noise or geometry descriptor.}
\label{fig:5}
\end{figure}

Geometry modified these profiles but did not yield the proposed monotone rule that a larger gap implies larger smoothing. The boundary-dominant design had the largest noisy mean smoothing (0.0861), while the more failure-prone clustered-gap design averaged 0.0661, and the marginal run-level correlations of $\log_{10}(\lambda)$ with fill distance, gap ratio, and nearest-neighbour CV were \ensuremath{-}0.035, \ensuremath{-}0.080, and \ensuremath{-}0.091, weak associations whose signs do not support the directional form of H2b. The combined standardized linear model reported $R^2$ = 0.4259, and a decomposition on the same output gave $R^2$ = 0.4096 for noise descriptors alone, $R^2$ = 0.0146 for geometry alone, and an increment of only 0.0163 when geometry was added to noise. Aggregating to the 240 slice tasks raised all values but preserved the conclusion, with a noise-only $R^2$ = 0.6573, a combined $R^2$ = 0.6835, and an incremental geometry contribution of 0.0262. Geometry therefore carries a detectable but modest smoothing-selection signal, and its stronger role lies in explaining failure severity rather than the scalar value of \ensuremath{\lambda}, which is the subject of \Cref{sec:3.4}. \Cref{fig:4,fig:5} display these relationships.

\subsection{Geometry contributes to failure-risk modeling}
\label{sec:3.4}

The additional regression analysis changes the response from smoothing magnitude to the probability of an exact-RBF failure. On the method-valid domain, the supplied summary reports McFadden pseudo-$R^2$ values of 0.180 for noise alone, 0.072 for geometry alone, and 0.266 for the combined model; the corresponding AUC values are 0.778, 0.656, and 0.828 \citep{hosmer2013,mcfadden1974,fawcett2006}. The geometry increment is reported as approximately 0.085 using unrounded estimates. This complements the small ordinary-$R^2$ increment for smoothing magnitude, but the two increments are not directly comparable effect sizes because they use different responses and different goodness-of-fit definitions.

The supplied joint Wald test of six geometry coefficients gives $\chi^2$ = 66.2 and p \ensuremath{\approx} 2.4\ensuremath{\times}$10^{-12}$ under slice-task-clustered standard errors \citep{cameron2015}. At the slice-task level, an ordinary least-squares model of the failure fraction increases from $R^2$ = 0.344 with noise alone to 0.484 with geometry and noise, an increment of 0.140. These summaries support an association within the benchmark, subject to the repeated-support and model-specification limitations in \Cref{sec:2.6}. They do not replace validation on independently generated geometries.

\begin{figure}[htbp]
\centering
\includegraphics[width=\linewidth,height=0.77\textheight,keepaspectratio]{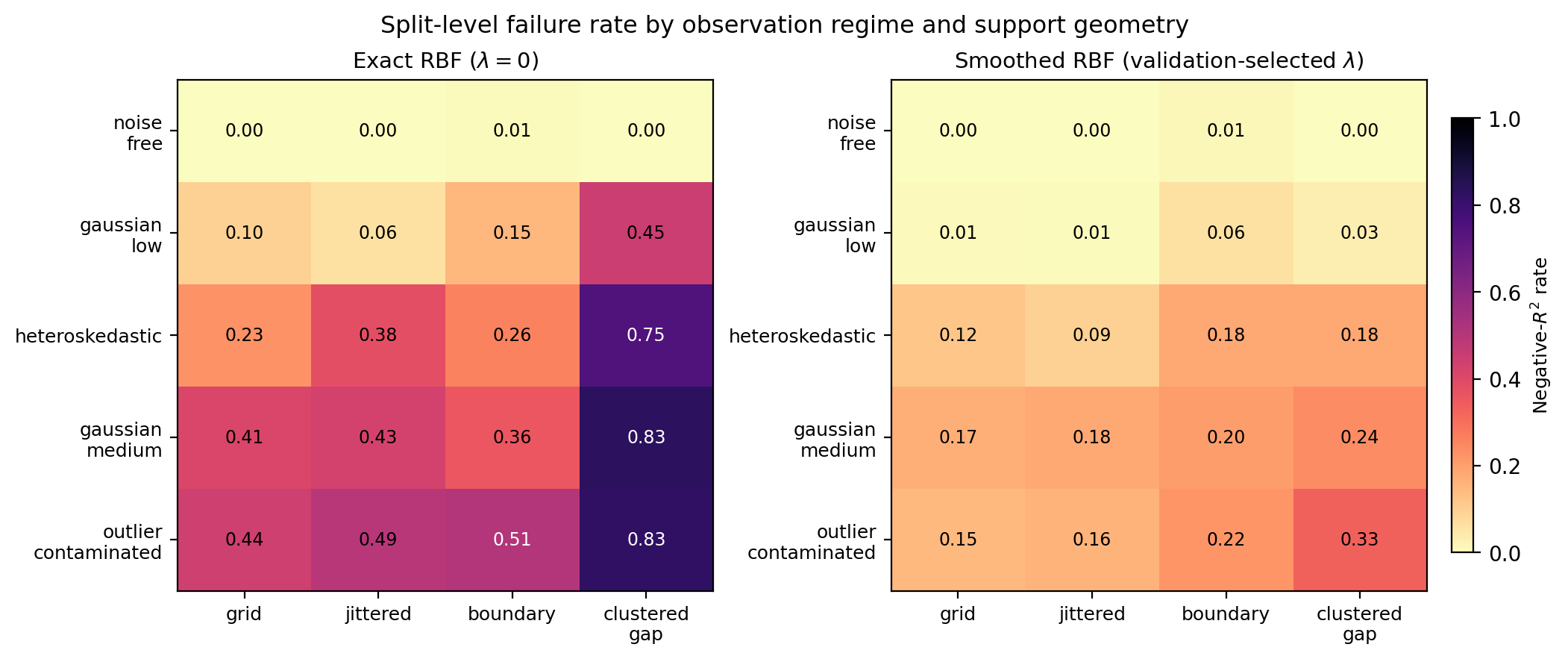}
\caption{Split-level failure rate (fraction of holdouts with $R^2$ < 0) by observation regime and support geometry. Exact RBF (left) fails with increasing frequency as both noise and geometric difficulty rise, peaking at the clustered-gap support; validation-selected smoothing (right) suppresses the pattern. The near-zero noise-free row shows that geometry sets vulnerability while noise activates it.}
\label{fig:6}
\end{figure}

\Cref{fig:6} shows that the geometry contrast depends on the observation regime. Exact-RBF failure is near zero for all geometries without noise, while noisy regimes exhibit substantial differences. Under medium Gaussian noise, the method-valid failure rates are 83.3\% for clustered-gap support, 41.3\% for the grid, and 35.8\% for boundary-dominant support. These are method-valid rates and should not be substituted for the shared-valid rates in \Cref{tab:4}. Smoothing attenuates the contrast, although residual failures remain. The pattern is consistent with a noise--geometry interaction; a formal interaction model is a separate analysis not established by the main-effects summaries alone.

Two qualifications constrain the interpretation. First, the reported geometry increment in McFadden pseudo-$R^2$ decreases from approximately 0.085 on the method-valid domain to 0.019 on the shared-valid domain. This is consistent with greater geometric sensitivity outside the common interior, but the restriction also changes the test points and available response variation, so the decrease does not by itself isolate a mechanism. Second, the descriptors are strongly correlated by construction: the clustered-gap design combines large fill distance, gap ratio, and anisotropy. Although the geometry block is jointly significant in the supplied analysis, only anisotropy is reported as marginally significant individually, at approximately p = 0.05. Individual coefficients should therefore not be interpreted as separable geometric effects. Direct conditioning diagnostics offer a useful complementary test, but are not interchangeable with coverage descriptors \citep{schaback1995}.

\begin{table}[htbp]
\centering
\small
\caption{Conductance-defined WQP slices after complete-case filtering. Conductance is expressed in the harmonized units used by the pilot (\(\mu\mathrm{S}/\mathrm{cm}\)); sites may occur in more than one slice.}
\label{tab:5}
\begin{tabularx}{\textwidth}{@{}Xrrrr@{}}
\toprule
Slice & Activities & Sites & Conductance range & Median \\
\midrule
Low & 655 & 21 & 25.6--313 & 230 \\
Mid & 646 & 17 & 314--467 & 376 \\
High & 649 & 16 & 468--22,200 & 1,600 \\
\bottomrule
\end{tabularx}
\end{table}

\subsection{Water Quality Portal reconstruction}
\label{sec:3.5}

The environmental pilot reproduced the broad robustness advantage of smoothing (\Cref{tab:6}). Averaged equally across conductance slices, smoothed RBF achieved RMSE 0.9877, MAE 0.6172, and mean $R^2$ = 0.8195 with no negative-$R^2$ split, cubic interpolation ranked second with RMSE 1.3031 and a negative-$R^2$ rate of 0.0417, and exact RBF was numerically unstable, with a mean RMSE of order $10^{31}$.

\begin{table}[htbp]
\centering
\small
\caption{WQP performance averaged across the three conductance slices. The exact-RBF mean is shown in scientific notation because the method fails on essentially every finite fit.}
\label{tab:6}
\begin{tabularx}{\textwidth}{@{}Xrrrrr@{}}
\toprule

Method & RMSE & MAE & Mean $R^2$ & Neg-$R^2$ rate & Valid fraction \\

\midrule

Cubic & 1.3031 & 0.7427 & 0.6531 & 0.0417 & 0.9762 \\

Exact RBF & 9.96\ensuremath{\times}$10^{30}$ & 7.71\ensuremath{\times}$10^{30}$ & \ensuremath{-}1.33\ensuremath{\times}$10^{63}$ & 0.6833 & 0.6833 \\

Smoothed RBF & 0.9877 & 0.6172 & 0.8195 & 0.0000 & 1.0000 \\

\bottomrule
\end{tabularx}
\end{table}

Exact-RBF failure was not confined to a few exceptionally large errors. The supplied results indicate that only 68.3\% of splits produced finite predictions and that every finite fit had negative $R^2$; the remaining splits returned no valid prediction. Thus, the tabulated negative-$R^2$ rate of 0.6833 is an unconditional fraction over all attempted splits, not evidence that the other 31.7\% succeeded. The source reports median finite-fit $R^2$ values ranging approximately from \ensuremath{-}$10^{17}$ to \ensuremath{-}$10^{26}$ across slices, with even the best finite fit below \ensuremath{-}170.

The environmental predictor space contains two distinct difficulties. Temperature and pH have very different raw scales, and the source reports covariance anisotropy ratios from approximately 285 to 980. Such anisotropy can affect an isotropic distance-based fit but does not by itself establish singularity. More decisively, between 194 and 301 activities per slice belong to duplicated temperature--pH pairs, with 48--69 pairs having different dissolved-oxygen values. Conflicting values at identical coordinates cannot be reproduced by a single-valued exact interpolant. Near-duplicates and the chosen kernel scale can additionally cause numerical ill-conditioning \citep{schaback1995,fornberg2004}. These observations make conditioning and coordinate treatment plausible explanations for instability, although the individual contributions were not isolated experimentally.

The within-coordinate diagnostic was recomputed from the supplied complete-activity data, grouping exactly equal stored temperature--pH pairs separately within each conductance slice. With $N$ activities and $G$ distinct coordinate groups, including singletons, the pooled estimator is $\widehat\sigma_{\mathrm{within}}^2=\sum_g\sum_i(y_{gi}-\bar y_g)^2/(N-G)$. Singleton groups contribute neither squared deviations nor degrees of freedom; their individual sample variance is undefined rather than zero. The resulting variances are 0.144645, 0.303271, and 0.148719 (mg/L)\textsuperscript{2} for the low-, mid-, and high-conductance slices, respectively. Dividing by the corresponding total slice variance, computed with denominator $N$, gives ratios of 2.72\%, 5.66\%, and 2.23\%. This reproduces the rounded range reported in the revised source using an explicit estimator; it does not establish the provenance of the original calculation. Under a representative, additive, homoscedastic replicate model, one minus these ratios would suggest an indicative $R^2$ ceiling of approximately 0.94--0.98. This interpretation remains conditional, not a demonstrated population bound: matching temperature and pH does not control monitoring site, date, conductance within a slice, or other environmental variables \citep{cressie1993}. Activity memberships, squared deviations, group counts, and denominators are supplied in the audit outputs described in \Cref{sec:audit}.

Every smoothed-RBF fit selected \ensuremath{\lambda} = 0.50, the upper candidate boundary. This establishes boundary saturation, but neither proves that the validation curve is monotone nor diagnoses over-regularization. The difference between achieved $R^2$ and the conditional replicate-based ceiling could reflect approximation bias, unmodeled variability, numerical instability, or limitations of the diagnostic itself. With no variation in the selected \ensuremath{\lambda}, a geometry--smoothing correlation is undefined, leaving H2b untested in this pilot. The next comparison should standardize or whiten the predictors, handle duplicate coordinates explicitly, and jointly tune scale and smoothing over a range broad enough to reveal an interior optimum if one exists \citep{fasshauer2007,fornberg2004}.

Across resamples, the mean two-coordinate optimum dispersion was 3.2117 for smoothed RBF, 7.0047 for cubic, and 12.2878 for exact RBF, which is directionally consistent with improved decision stability but should not be read as a dimensionless physical effect, because temperature and pH were combined without scale standardization and no true environmental optimum was available for validation.

\section{Discussion}
\label{sec:4}

\subsection{What the benchmark establishes}
\label{sec:4.1}

The strongest result is the conditional failure of exact interpolation. When observations were clean, exact multiquadric RBF reconstruction was highly accurate, but once noise was introduced, exact fitting increased both mean error and the probability of split-level collapse, and smoothing recovered much of the lost performance while improving tail-oriented diagnostics. This supports H1 and is consistent with the role of the diagonal penalty $\lambda I$ in reducing sensitivity to perturbations in the interpolation system, the kernel expression of the bias--variance trade-off that motivates ridge-type estimators \citep{hoerl1970,hastie2009}. Geometry mattered most as a vulnerability descriptor, since the clustered-gap design, which combined short within-cluster separations with a large, uncovered region and strong anisotropy, produced by far the worst exact-RBF behaviour under noise yet remained benign in the clean benchmark. This supports the failure-oriented H2a, in which coverage and site distribution determine how measurement perturbations are amplified \citep{schaback1995}.

\subsection{Geometry has different roles in failure and smoothing}
\label{sec:4.2}

The two forms of H2 lead to different conclusions. Geometry contributes little to the fitted model of scalar smoothing, but the supplied failure analysis shows an additional association with exact-RBF instability. This does not invalidate smoothing magnitude as a research question; it indicates that failure probability and selected \ensuremath{\lambda} capture different aspects of reconstruction difficulty. A single global \ensuremath{\lambda} must accommodate both dense and sparse regions, the shape heuristic may absorb some geometric effects, the candidate grid is discrete, and 16-point tasks provide limited resolution for correlated support descriptors \citep{fasshauer2007,fornberg2004}.

Conditioning is a candidate mechanism rather than an established mediator. A future analysis should log a clearly specified condition-number diagnostic for the actual augmented RBF system, with and without regularization, and relate it to support geometry and failure. Because multiquadric reconstruction includes a polynomial constraint, the relevant system is not simply an unconstrained K matrix. Moreover, matrix conditioning does not encode all coverage or extrapolation effects. Condition numbers should therefore complement, not automatically replace, geometric descriptors. Testing whether they account for the observed association requires controlled rescaling, duplicate handling, and independent evaluation \citep{schaback1995,demarchi2010}.

\subsection{Why the WQP pilot is informative despite weak H2b evidence}
\label{sec:4.3}

The WQP study confirms that environmental feature-space data can violate the assumptions of exact interpolation more severely than designed synthetic data. Repeated temperature--pH coordinates with different dissolved-oxygen responses are not anomalous in a monitoring database, because unmodelled variables, seasonality, site differences, laboratory variation, and rounding all produce distinct outcomes at nominally identical coordinates, so the data do not define a single exact function of temperature and pH and regularization changes an inconsistent interpolation problem into an approximation problem. At larger scales, deep-learning models have predicted dissolved oxygen across many basins, including basins without local observations, showing that transfer across sites is feasible but sensitive to how site heterogeneity is represented \citep{zhi2021}, which reinforces that treating dissolved oxygen as an exact function of two predictors is not physically justified. At the same time, the pilot cannot validate geometry-controlled smoothing, because the three conductance strata come from one county-level dataset and share collection processes, there are only three coarse slice classes, coordinates are unstandardized, duplicate points are common, and all fits saturate at the same boundary, which leaves essentially no independent smoothing variation to explain; the weak field result is therefore a limitation of design and parameter range rather than evidence that geometry is irrelevant.

\subsection{Implications for geometry-aware model selection}
\label{sec:4.4}

A geometry-aware workflow should first check whether exact fitting is admissible by identifying duplicate coordinates, conflicting responses, and rank deficiencies. It should then assess support quality through gaps, separation, anisotropy, hull coverage, and boundary concentration, together with a conditioning diagnostic for the fitted system. These checks can flag risk but should not be presented as a validated classifier until tested on independent support patterns. Finally, the RBF scale and smoothing should be tuned jointly, expanding the candidate range when the selected value remains at a boundary. Local reconstruction or adaptive kernels provide alternatives when one global penalty cannot accommodate heterogeneous support \citep{hesse2021,wenzel2021,demarchi2020,bozzini2015}. This workflow separates admissibility, geometric risk, and parameter selection rather than assuming that one descriptor determines all three.

\subsection{Limitations}
\label{sec:4.5}

Several limitations constrain interpretation. The synthetic benchmark uses small 16-node slices and four designed geometry classes, the descriptors are correlated by construction, and repeated holdouts are not independent replicates of geometry, so the exploratory linear model is descriptive rather than inferential \citep{hurlbert1984}. The RBF shape parameter was set by a median-distance heuristic, and because multiquadric conditioning depends jointly on scale, smoothing, and geometry, fixing \ensuremath{\varepsilon} may attribute some scale effects to \ensuremath{\lambda}, so joint nested tuning is needed \citep{fasshauer2007,fornberg2004}. The smoothing grid is bounded and discrete, and the upper-bound saturation in every WQP split prevents correlation analysis and implies that stronger candidate values must be considered, while lower-bound selections in the clean benchmark approximate exact interpolation without including \ensuremath{\lambda} = 0 in the validation grid. The boundary-dominant synthetic implementation retains only perimeter nodes, which is a useful stress case that should not be generalized to every boundary-biased program, so future experiments should vary the interior-to-boundary ratio continuously. The WQP reconstruction uses temperature and pH in original units, so Euclidean distances, fill distance, anisotropy, optimum dispersion, and the median-distance scale are unit-dependent and should be standardized or transformed under a justified metric before a definitive geometry analysis. Random activity-level splitting allows the same monitoring site, date neighbourhood, or repeated coordinate to appear in both training and test sets, so grouped temporal and site-held-out validation would provide a stricter assessment of environmental generalization \citep{roberts2017}. Finally, the environmental surface is descriptive rather than causal, since dissolved oxygen depends on variables beyond temperature, pH, and conductance, and the fitted surface should not be read as a mechanistic oxygen model.

\subsection{Next experimental phase}
\label{sec:4.6}

The next study should use multiple counties, watersheds, or agency programs as independent geometry units, standardize the predictors within each unit, aggregate exact duplicate coordinates while retaining within-coordinate variance as a noise descriptor, and hold out monitoring sites or time blocks during validation to respect the dependence structure of environmental data \citep{roberts2017}. The RBF scale and smoothing should be jointly tuned over an expanded grid with nested grouped cross-validation \citep{golub1979,fasshauer2007}, and the condition number of the kernel system should be logged alongside global and local geometry descriptors \citep{schaback1995}. A decisive test of H2 would compare nested predictive models for selected regularization or failure risk, spanning noise-only, geometry-only, noise-plus-geometry, and interaction models, evaluated on entirely held-out basins or programs, so as to determine whether geometry provides transferable information rather than merely describing the support patterns used to build a benchmark.

\section{Conclusion}
\label{sec:5}

The synthetic benchmark supports regularization under noisy observations and identifies clustered-gap support as a particularly difficult configuration for exact multiquadric interpolation. The distinction between H2a and H2b is essential: geometric descriptors are associated with failure risk within the tested designs, but contribute comparatively little to the variation in validation-selected smoothing after noise is considered. The reported failure-regression results strengthen the case for further screening research, subject to their dependence structure and the need for independent validation.

The environmental pilot confirms that repeated predictor coordinates with conflicting responses are incompatible with exact single-valued interpolation. Smoothed RBF gives substantially more reliable predictions, but universal upper-bound selection prevents a geometry--smoothing analysis. Unequal coordinate scales and possible ill-conditioning motivate corrective experiments; they do not, by themselves, demonstrate over-regularization or a causal pathway to failure. The combined conclusion is therefore that geometry-aware failure screening is supported as an exploratory direction, whereas a transferable geometry-driven smoothing policy remains unvalidated. The next stage requires independent field geometries, appropriate scaling, explicit duplicate handling, joint parameter tuning, and reproducible conditioning and noise diagnostics.

\section*{Data and Code Availability}

The supplied project contains the original experiment scripts in \path{MATH-PAPER/experiments/}, 17,280 synthetic run-level records in \path{MATH-PAPER/outputs/runs.csv}, and 360 environmental run-level records in \path{MATH-PAPER/outputs_real/runs_real.csv}. The cached county extract and 1,950 complete activities are available under \path{MATH-PAPER/data/raw/} and \path{MATH-PAPER/outputs_real/}, respectively. A local audit script, \path{MATH-PAPER/experiments/audit_reproducibility.py}, reproduces the smoothing-model summaries and within-coordinate diagnostic without downloading data or overwriting the original results. Its outputs, source checksums, reconstructed task identifiers, group memberships, and audit-environment metadata are in \path{MATH-PAPER/reproducibility_audit/}; \Cref{sec:audit} specifies their interpretation and limitations.

The WQP query requested January 1, 2020--December 31, 2025. The cached raw county records span January 6, 2020--August 5, 2025, whereas the complete activities span January 6, 2020--December 10, 2024. The original retrieval timestamp and a public data/code repository identifier were not found and remain to be supplied; file-upload dates are not retrieval dates. The supplied \path{MATH-PAPER/requirements.txt} contains minimum dependency versions rather than a pinned original environment. The separate six-predictor logistic/Wald script and its fitted outputs, candidate-level validation logs, and condition-number records were not present. Consequently, the reported logistic, clustered-Wald, and failure-fraction regression statistics remain unverified, even though the underlying run-level outcomes are now available.

\section*{Acknowledgments}

The author acknowledges the organizations contributing observations to the Water Quality Portal. Additional funding, institutional, and contributor statements can be inserted before submission.

\bibliographystyle{unsrtnat}

\appendix
\section{Reproducibility audit of the supplied artifacts}
\label{sec:audit}

The September 5, 2026 audit distinguishes archived experiment outputs from newly reconstructed diagnostics. It does not rerun interpolation, infer an undocumented six-predictor model, or replace the original files. The appendix is included directly in this manuscript so that compilation does not depend on a separate audit file. The supplemental audit script and generated output directory described below are not present in the current project snapshot and must be restored before a complete reproducibility release.

\subsection{Predictor definitions and recoverable fitting specification}

The supplied synthetic code defines nine geometry predictors: mean, coefficient of variation, minimum, and maximum nearest-neighbour distance; empirical fill distance; gap ratio; convex-hull area ratio; boundary fraction; and covariance anisotropy ratio. These are computed on each complete 16-node task and reused across its holdouts. They are not a recovered specification of the six-coefficient geometry block in the reported logistic analysis.

For the complete task residuals $e_i=y_i-f(\bm{x}_i)$, the four available noise predictors are
\begin{align}
 s_e &= \left[n^{-1}\sum_i(e_i-\bar e)^2\right]^{1/2}, &
 a_e &= n^{-1}\sum_i|e_i|,\\
 o_e &= n^{-1}\sum_i\mathbf{1}\{|e_i|>2.5s_{\mathrm{rob}}\}, &
 r_e &= \left[n^{-1}\sum_i e_i^2\right]^{1/2}.
\end{align}
Here $s_{\mathrm{rob}}=1.4826\,\operatorname{median}_i|e_i-\operatorname{median}_j(e_j)|$ when the median absolute deviation is positive, and $s_{\mathrm{rob}}=\max(s_e,10^{-12})$ otherwise. Their stored names are \texttt{empirical\_noise\_std}, \texttt{empirical\_noise\_mae}, \texttt{outlier\_fraction}, and \texttt{noise\_scale\_proxy}. The supplied code does not identify which of these entered the separate logistic regression.

The recoverable smoothing analysis uses $\log_{10}(\max(\lambda,10^{-8}))$ as the response, centers each predictor, divides by its population standard deviation (substituting 1 for a zero standard deviation), adds an intercept, and fits ordinary least squares using \texttt{numpy.linalg.lstsq} with \texttt{rcond=None}. The earlier audit recalculation from the saved runs gave geometry-only, noise-only, and combined $R^2$ values of 0.014619, 0.409609, and 0.425944. Averaging the log-response over the 24 holdouts in each task gave the corresponding task-level values 0.023457, 0.657273, and 0.683485, reproducing the smoothing results in \Cref{sec:3.3}. That audit also regenerated task observations with seed 42; all nine geometry and four noise descriptors agreed with the archived values within relative tolerance $10^{-9}$ and absolute tolerance $10^{-10}$. Regenerated observations must be distinguished from archived data.

\subsection{Identifiers, failure outcomes, and unresolved inference}

The saved fields identify 240 tasks by the tuple (family, geometry, noise regime, slice level), with 24 repeated holdouts and three methods per task. Removing noise regime gives 48 design-group labels; these are not 48 independent support configurations because deterministic designs recur. These reconstructed labels cannot be certified as the labels used by the original clustered-covariance implementation. That implementation, its exact six geometry variables, noise block, filtering, fitted coefficients, fitted probabilities, covariance corrections, and test conventions were not found.

The saved data contain 5,760 exact-RBF run records from which 240 task-level failure summaries can be reconstructed. The archived shared-domain summary computes the fraction of \emph{all} runs satisfying $R^2<0$, leaving undefined values in the denominator with no failure flag. This is not the failure rate conditional on a defined shared $R^2$. Both conventions must be reported with their denominators; undefined values must not be interpreted as successful reconstruction. The tabulated shared-domain fractions retain the original convention. The earlier audit reproduced the observed method-valid medium-Gaussian failure fractions, but the logistic pseudo-$R^2$, AUC, clustered Wald test, shared-domain regression sensitivity, and task-level failure-regression coefficients cannot be verified without the missing model specification. No guessed replacement model is presented as the original analysis.

\subsection{Within-coordinate variance and data membership}

The earlier audit recreated all 1,950 complete activities from the cached county CSV using the supplied preprocessing. Within each conductance stratum, exactly equal stored temperature and pH values define a group, without additional rounding or near-duplicate merging. Let $n_g$ be a group size, $G$ the number of groups including singletons, and $N=\sum_g n_g$. The explicit estimator is
\begin{equation}
 \widehat\sigma_{\mathrm{within}}^2
 =\frac{\mathrm{SSE}_{\mathrm{within}}}{N-G},\qquad
 \mathrm{SSE}_{\mathrm{within}}=\sum_{g=1}^{G}\sum_{i=1}^{n_g}(y_{gi}-\bar y_g)^2.
\end{equation}
Singletons contribute zero to both numerator and denominator, while their individual sample variance is undefined. Total slice variance uses denominator $N$. The squared-deviation numerator, effective denominator, and resulting estimate are shown in \Cref{tab:audit-variance}.

\begin{table}[htbp]
\centering
\small
\caption{Recomputed within-coordinate diagnostics. Variances are in (mg/L)$^2$. $G$ includes singleton groups.}
\label{tab:audit-variance}
\begin{tabular}{@{}lrrrrr@{}}
\toprule
Slice & $N$ & $G$ & $N-G$ & Within-group SSE & Pooled variance\\
\midrule
Low & 655 & 533 & 122 & 17.646667 & 0.144645\\
Mid & 646 & 509 & 137 & 41.548119 & 0.303271\\
High & 649 & 455 & 194 & 28.851500 & 0.148719\\
\bottomrule
\end{tabular}
\end{table}

The low, mid, and high strata contain 461, 428, and 348 singleton groups, respectively. The audit generated \path{within_coordinate_members.csv}, \path{within_coordinate_groups.csv}, and \path{within_coordinate_variance.csv} to record activity identifiers, group membership, group means, squared deviations, counts, and variance ratios. These supplemental files must accompany the reproducibility release; the underlying complete-activity CSV remains available in the project. Coordinate replicates do not isolate measurement noise from unmodeled environmental variation.

\subsection{Validation logging, geometry implementation, and provenance}

The supplied inner-validation function evaluates the eight candidates in \Cref{sec:2.2} over five randomized holdouts, averaging RMSE over successful finite evaluations. Candidates without a valid score are omitted; if none is valid, the function returns 0.01. The shape parameter is recalculated from each fitted subset, including inner-training subsets. Only the selected smoothing is retained in the run CSVs. All 120 environmental smoothed-RBF runs select 0.50, but their candidate-level scores, successful-fold counts, split memberships, and condition numbers were not saved. An instrumented rerun would therefore generate new diagnostics rather than recover original logs.

The code inspection also identifies a coverage-metric limitation beyond unequal coordinate scaling. The environmental script reuses the synthetic geometry routine unchanged: its fill-distance reference grid and boundary tests remain tied to $[1,2]\times[0.5,1.5]$, and its hull measure is unnormalized area. The stored environmental fill distance, boundary fraction, and hull-area-ratio label therefore do not describe coverage of an appropriately defined temperature--pH domain. They require an explicitly corrected experimental rerun and should not be used as evidence of an environmental geometry--smoothing mechanism. The original interpolation outputs have not been modified.

The original dependency file specifies NumPy $\geq2.0$, pandas $\geq2.2$, SciPy $\geq1.14$, and Matplotlib $\geq3.9$, without pinned versions. The earlier \emph{audit}, not the original experiment, ran with Python 3.13.15, NumPy 2.4.1, pandas 2.3.3, SciPy 1.17.0, and Matplotlib 3.10.8. The exact query string is retained in the supplied environmental experiment script. The original retrieval date, public repository identifier, and original locked environment remain unavailable. An additional statewide ZIP file in the upload could not be opened as a complete archive during the audit and was not used in the county-level calculations.

\bibliography{sancak_revised_references}

\end{document}